\documentclass[11pt,letterpaper]{llncs}
\let\llncssubparagraph\subparagraph
\let\subparagraph\paragraph
\usepackage[compact]{titlesec}
\let\subparagraph\llncssubparagraph

\usepackage[font=scriptsize,labelfont=bf,skip=1pt]{caption}
\everypar{\looseness=-1 }
\usepackage{newtxtext}
\usepackage{fancyhdr}

\usepackage{multirow}
\usepackage{booktabs}
\usepackage{graphicx} 
\usepackage{amssymb}
\usepackage{amsmath}
\usepackage{amssymb}
\usepackage{graphicx}
\usepackage{bmpsize}
\usepackage{url}
\usepackage{placeins}
\usepackage{mathtools}
\usepackage{enumitem}

\usepackage{geometry}
\newcommand{\keywords}[1]{\par\addvspace\baselineskip
\noindent\keywordname\enspace\ignorespaces#1}

\usepackage{epstopdf}
\usepackage{setspace}
\usepackage{subcaption}
\usepackage{wrapfig}
\DeclareGraphicsExtensions{.eps}

\usepackage[open,openlevel=1]{bookmark}
\usepackage[dvipsnames,table]{xcolor}
\usepackage{tablefootnote}

\usepackage{hyperref}
\hypersetup{
	colorlinks   = true,    
	urlcolor     = blue,    
	linkcolor    = blue,    
	citecolor    =  blue,      
	bookmarksopen=true
}

\fancypagestyle{firstpage}{\fancyhf{}
\setcounter{page}{1}
\fancyhead[C]{\scriptsize{International Journal of Data Mining \& Knowledge Management Process (IJDKP) Vol.7$,$ No.5$/$6$,$ November 2017}}
\fancyfoot[L]{DOI: 10.5121/ijdkp.2017.7601}
\rfoot{\thepage}
}

\begin{document} 
	\title{\LARGE{On Using Network Science in Mining Developers Collaboration in Software Engineering: A Systematic Literature Review}}

\author{\large{Mohammed Abufouda \and Hadil Abukwaik}}
\institute{\large{University of Kaiserslautern, Department of Computer Science,\\ Kaiserslautern 67663, Germany}\\ 
\email{\{abufouda,abukwaik\}@cs.uni-kl.de} }

\maketitle
\thispagestyle{firstpage}
\begin{abstract}		\textit{Background}: Network science is the set of mathematical frameworks, models, and measures that are used to understand a complex system modeled as a network composed of nodes and edges. The nodes of a network represent entities and the edges represent relationships between these entities. Network science has been used in many research works for mining human interaction during different phases of software engineering (SE).\\
		\textit{Objective}: The goal of this study is to identify, review, and analyze the published research works that used network analysis as a tool for understanding the human collaboration on different levels of software development. This study and its findings are expected to be of benefit for software engineering practitioners and researchers who are mining software repositories using tools from network science field.\\
		\textit{Method}: We conducted a systematic literature review, in which we analyzed a number of selected papers from different digital libraries based on inclusion and exclusion criteria. \\
		\textit{Results}: We identified $35$ primary studies (PSs) from \textit{four} digital libraries, then we extracted data from each PS according to a predefined data extraction sheet. The results of our data analysis showed that not all of the constructed networks used in the PSs were valid as the edges of these networks did not reflect a real relationship between the entities of the network. Additionally, the used measures in the PSs were in many cases not suitable for the used networks. Also, the reported analysis results by the PSs were not, in most cases, validated using any statistical model. Finally, many of the PSs did not provide lessons or guidelines for software practitioners that can improve the software engineering practices.\\
		\textit{Conclusion}: Although employing network analysis in mining developers' collaboration showed some satisfactory results in some of the PSs, the application of network analysis needs to be conducted more carefully. That is said, the constructed network should be representative and meaningful, the used measure needs to be suitable for the context, and the validation of the results should be considered. More and above, we state some research gaps, in which network science can be applied, with some pointers to recent advances that can be used to mine collaboration networks. 
\end{abstract}
	

	\section{Introduction}
	\label{sec:introduction}
	Network science is the set of mathematical frameworks, models, and measures that are used to analyze a complex system modeled as a network. Since the seminal works of Watts and Strogatz~\cite{watts1998} and Barab{\'a}si and Albert~\cite{barabasi1999}, the field of network science exploded in different directions with a huge amount of measures, models, and applications for network analysis. Fields like Biology, Social science, Physics, and Computer science contribute a lot to the network science field, each from different perspectives. Other domains benefited from the measures and tools provided in the field of network science, and software engineering is one of those fields. Software engineering is a complex production system, in which humans are involved in order to produce a software with some certain properties. One face of the complexity of software engineering process is the interaction among developers who develop a software. This complexity can be seen in many phases of the software engineering life cycle including collaborative coding, bug fixing and tracking, communication in mailing lists, and many other interactions. The interactions between developers can be seen as a \textit{network} where the nodes are, in most cases, the developers and the edges between these nodes are the interactions between them. Having these interaction constructed as networks opens the way to use network science as an effective tool to analyze these networks. In this work, we systematically review the application of network analysis for the networks constructed from the collaboration interaction between software developers.
	\subsection{Research Questions}
	\label{subsec:researchquestions}
	The course of improving software engineering process, practices, and software quality is hard and it get even harder when considering the human factor. To that end, a lot of work has(is) being done and a lot of venues have been dedicated for mining the human artifacts (like developers' collaboration and communication interactions) and the software artifacts (like the source code, the requirements and specification documentation) in order to gain insights to improve the software engineering.
	In this research, we aim at reviewing, evaluating, and providing future directions for the use of network science in the area of software engineering specifically for the collaboration between developers. Thus, this work has the following research questions:
	\begin{itemize}
		\item[$\circ$] \textbf{RQ1}: \textit{How valid and reliable are the constructed networks?}
		The main criterion for conducting good network analysis is to have a valid network, otherwise, any subsequent analysis may be useless. Thus, this research question investigates the validity and the quality of the constructed networks in the identified studies. Here, we explore the quality of the edges in the constructed networks, like being a real or a proxy relationship, being aggregated over time (longitudinal aggregation) and space (multiplex aggregation).
		\item[$\circ$] \textbf{RQ2}: \textit{How valid and reliable is the use of network analysis measures?}
		Network science is noticeably multidisciplinary, which renders some measures not suitable for every network and in every context. This research question investigates how meaningful the use of certain measures for the constructed networks is. We also shed light on the unstated assumptions of some of the measures that were used intensively in the primary studies.
		\item[$\circ$] \textbf{RQ3}: \textit{How valid, reliable, and generalizable are the results reported by the primary studies?}
		In order to get the confidence regarding the significance of the reported results by any primary study, a validation should be performed. This research question investigates the validation status of the reported results that were based on using network analysis. We also explore the generalizability of the reported results by examining the number of analyzed networks in the study.
		\item[$\circ$] \textbf{RQ4}: \textit{How are the results being reflected on the studied context?} Having data is tempting to start analysis, however, any analysis that does not provide actionable and useful insights for the studied context should be avoided. This research question investigates whether the PSs reflected the results on the studied context of software engineering or not. 
		\item[$\circ$] \textbf{RQ5}: \textit{What are the gaps in the research that can be further researched?}
		Based on our extracted data and our analysis results, we identify potential research directions to cover existing gaps about where network analysis can be employed. The answer of this question should help in extending the body of the research in mining developers networks using network analysis. Additionally, we provide some pointers to recent advances in the network science that can be utilized.
	\end{itemize}
	\subsection{Method}
	\label{subsec:method}
\begin{figure}
		\begin{center}
			\includegraphics[width=0.5\textwidth]{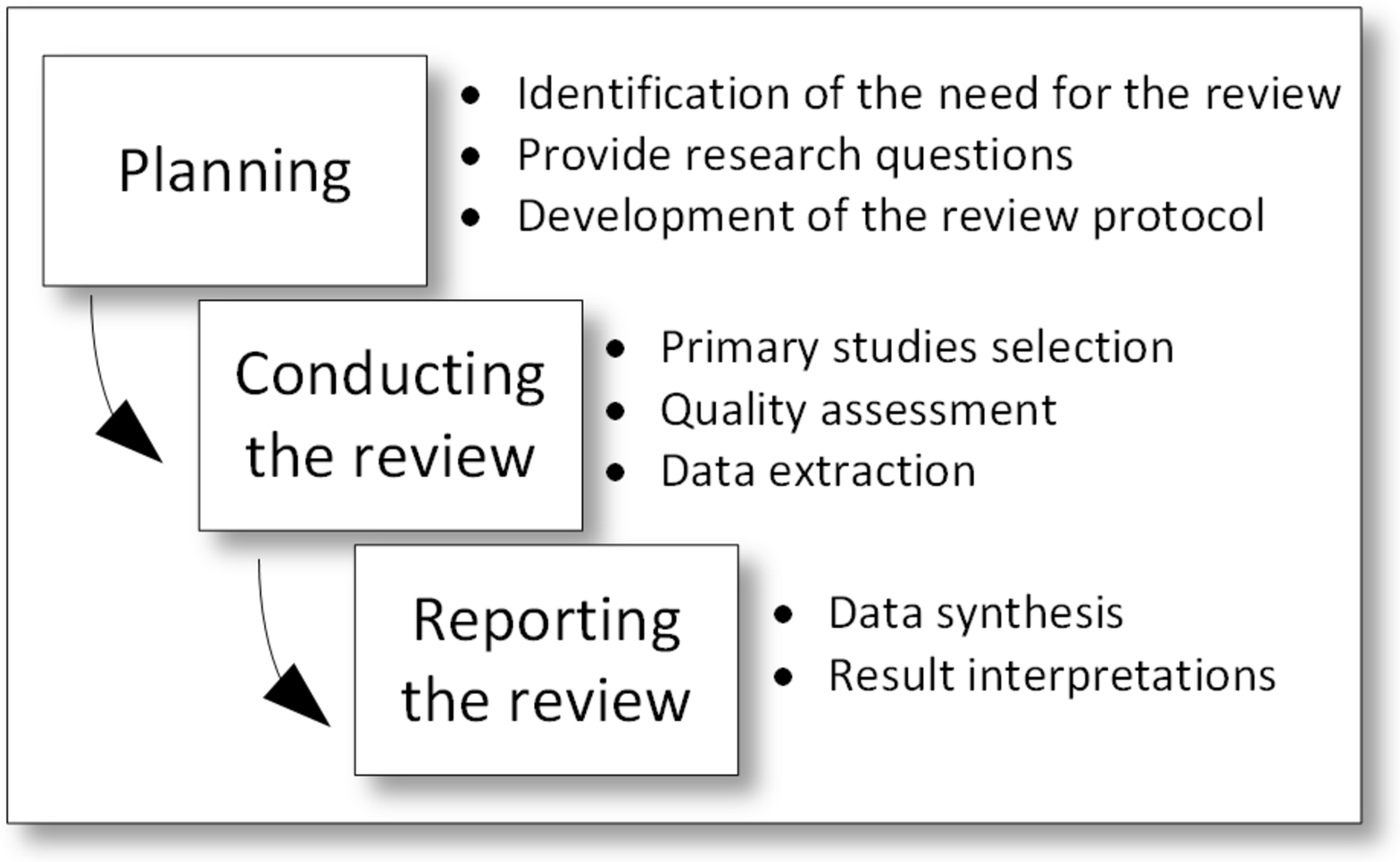}
		\end{center}
		\vspace*{-5mm}
		\caption{The systematic review steps(guidelines) adapted from Kitchenham and Charters~\cite{Kitchenham2007}}		
		\label{fig:reviewsteps}
	\end{figure}

\begin{figure*}
		\begin{center}
			\includegraphics[width=\textwidth]{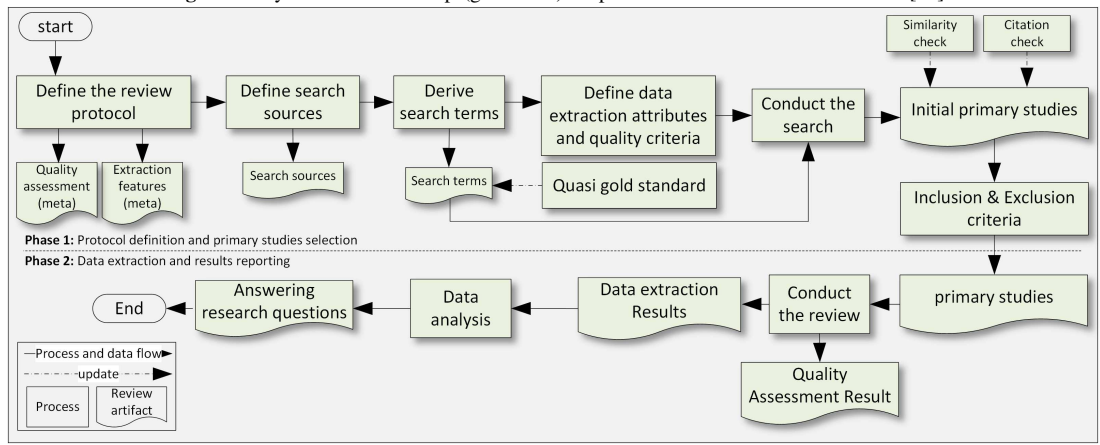}
		\end{center}
		\caption{The systematic review process model followed in this SLR.}		
		\label{fig:reviewprocessmodel}
	\end{figure*}

	In this systematic literature review (SLR), we followed the guidelines provided by Kitchenham and Charters~\cite{Kitchenham2007}. They offer the most formal and strict review process we know in empirical software engineering. Figure~\ref{fig:reviewsteps} abstracts the steps that we adapted from Kitchenham and Charters~\cite{Kitchenham2007} and implemented in this SLR. Based on these guidelines, we built our process model for conducting the review. Figure~\ref{fig:reviewprocessmodel} shows this model in detailed steps. In the following section, we provide details about the model.
	The rest of this paper is organized as follows. Section~\ref{sec:definitions} provides the definitions that will be used in this paper. Section~\ref{sec:theproceduresofthereview} gives details about the followed steps conducted in this SLR. Section~\ref{sec:results} contains the results of the extracted data, while Section~\ref{sec:discussion} provides the answers to the research questions. This paper concludes in Section~\ref{sec:conclusion}\footnote{The data of this SLR is available upon request.}.

	\section{Definitions}
	\label{sec:definitions}
	In this section, we provide the definitions~\footnote{We will stick to the terminology used in network science field to help researchers in software engineering in the future in finding the related work more easily.} required to proceed in this review. We also provide information about how network analysis should be properly applied in order to get meaningful results. The data extraction fields presented in Section~\ref{subsec:dataextraction} are based on the definitions in this section.\\
	An undirected network  $G=(V,E)$ is a tuple that is composed of two sets $V$ and $E$, where $V$ is the set of nodes and $E$ is the set of edges such that an undirected edge $e$ is defined as $e=\{u,v\} \in E$ where $u,v \in V$. For a directed network $\overrightarrow{G}=(V,\overrightarrow{E})$, a directed edge $\overrightarrow{e}$ is defined as $\overrightarrow{e}=(u,v)$ where the node $u$ is the \textit{source} and the node $v$ is the \textit{target}. For undirected networks, the \textit{Degree Centrality} of a node $w$, $deg(w)$, is defined as the number of nodes that are connected to it, while for directed networks the \textit{in-degree} and the \textit{out-degree} are defined as the number of edges in the network where the node $w$ is the target and the source node, respectively. The set of all neighbors of a node $v$ is denoted as $N(v)$. Graphs can be weighted, i.e., edges are associated with a weight that reflects the intensity, frequency, or the distance between two nodes. Thus, a weighted undirected network is defined as: $G=(V,E,\omega)$, where $\omega: E\rightarrow \mathbb{R}$. Networks whose nodes and edges are fixed over time are called \textit{static} networks. Networks that consider the temporality of nodes and edges are called \textit{dynamic} or \textit{temporal} networks, and normally denoted as $G_t(V,E)$, e.g, the network $G$ at time point $t$.
	
	\subsection{Centrality measures}
	\label{subsec:centralitymeasures}
	A set of measures is defined based on the definition of a network. The \textit{distance} (reachability) between two nodes $u$ and $v$ is defined as: $d(u,v)$ which is the shortest path between the two nodes. The \textit{diameter} of a network is the maximal distance between any two nodes in the network. The \textit{clustering coefficient} of a node $v$ is defined as: $cc(v)=\frac{e(v)}{deg(v)(deg(v)-1)/2}$ for directed networks and as: $cc(v)=\frac{e(v)}{deg(v)(deg(v)-1)}$ for undirected networks. The \textit{network density} measures how many edges there are in the network compared to the maximum possible number of edges, and it is defined as: $\eta (G) = \frac{2m}{n(n-1)}$ for undirected networks, where $m = |E|$ and $n = |V|$. The networks with high density are called \textit{dense} networks and those with low density are called \textit{sparse} networks. The \textit{betweenness centrality} of a node $v$ is defined as: $B(v)=\sum_{s\in V(G)}\sum_{t\in V(G)}\frac{\sigma_{st}(v)}{\sigma_{st}}$, where $\sigma_{st}(v)$ is the number of shortest paths between the nodes $s$ and $t$ that includes the node $v$, while $\sigma_{st}$ is the number of all shortest paths between the nodes $s$ and $t$. The \textit{closeness centrality} of a node $v$ is defined as: $C(v)=(\sum_{w\in V(G)}d(v,w))^{-1}$.\\
	\subsection{Bipartite networks}
	\label{subsec:bipartite}
	A special case of networks is the \textit{Bipartite} networks (2-mode networks or affiliation networks), which is defined as $G(V_L,V_R,E)$, where $V = V_L\cup V_R$, $V_L \cap V_R = \phi$, and $E: V_L \rightarrow V_R$. The bipartite networks are appropriate model when we want to model the interactions between two disjoint entities, like developer and source code. A method called \textit{One Mode Projection} (OMP) is classically used to convert the bipartite networks into a unipartite network, where the nodes in the projected network consists of only one type of nodes.\\
	\subsection{Multiplex}
	\label{subsec:multiplex}
	Networks can be \textit{Multiplex} (Multi-layer, multilateral, etc.), where the interactions between the nodes have many types, each can be represented as a different network. The formal definition of this type of networks is: $\mathcal{M}= (\mathcal{G},\mathcal{C})$ where $\mathcal{G}= \{{G_{\alpha}; \alpha \in {1,\cdots,M}}\}$ is the set of all layers, each of them $G_{\alpha} \in \mathcal{G}$, $G_{\alpha}= (V_{\alpha},E_{\alpha})$ represents one type $\alpha$ of interactions. The inter-layer edge $\mathcal{C}$ is defined as: $\mathcal{C}=\{E_{\alpha\beta} \subseteq V_{\alpha}   \times V_{\beta};\ \alpha,\beta \in \{1,\cdots,M\},\ \alpha \neq \beta\}$ represents the number of edges between the nodes from two different layers. Normally, when $\mathcal{C} = \phi$, the networks are called multiplex, otherwise, it is called multilayer.
	\subsection{Null models and network motifs}
	\label{subsec:motifs}
	\textit{Null models} are graph models that are used to test the statistical significance of the results obtained from an observed (constructed) network. Using a null model gives more confidence and creditability to the conclusions of any network-based measures or models. \textit{Random Graphs} with the same degree sequence of the observed network are wildly used null model.
	Network \textit{motif}~\cite{shen2002} is a subgraph that exists significantly higher than the corresponding null model. Finding those motifs gives insights about the patterns that occur frequently in a network.
	
	\section{The procedures of the review}
	\label{sec:theproceduresofthereview}
	In this section, we provide a full description of the SLR process, which we followed in this  study. This process is illustrated in Figure~\ref{fig:reviewprocessmodel}. We started by defining the goal of the review, which is a row version of the presented research questions in Section~\ref{subsec:researchquestions}. Accordingly, we defined the \textit{Quality Assessment} (QA) items of the PSs and the \textit{Extraction Features} (EF) items for the analysis of the PSs. The search sources were then defined and the \textit{search terms} where identified. To add more strictness to the review, we followed the \textit{Quasi-Gold Standard} method presented by Zhang et al.~\cite{zhang2010searching} to select the appropriate \textit{search terms} semi-systematically as full-systematic term identification is hard to achieve and to reproduce. Having our search terms identified, the search process was conducted on the defined search sources producing the \textit{raw primary studies}. The raw primary studies were checked by snowballing , which is a forward and backward citation check. We did also a similarity check to find similar papers for each paper in the primary studies. The similarity check is provided in some libraries' search engine that enabled a library user to see similar and related papers. Having the raw primary studies, we applied the inclusion and exclusion criteria to get the final set of the \textit{primary studies}. 
	
	\subsection{Quasi-Gold Standard search}
	\label{subsec:QGS}
	One challenge in performing a rigor SLR is to identify the search terms systematically. To the best of our knowledge, the most rigorous method to attain high reproducibility is the Quasi-Gold Standard (QGS) provided by Zhang et al.~\cite{zhang2010searching}. This method starts with manually identifying a set of studies that are relevant to the SLR and a set of search terms. Then, the researchers run the automatic search in the search libraries many times, in each run then, the quality of the retrieved papers is evaluated. The evaluation is done by calculating the \textit{sensitivity} of the search result (i.e., the number of related studies found by the search divided by the number of QGS studies). In each run, the search terms are updated by introducing new relevant terms until we reach the targeted sensitivity, which should be $100\%$ in the best situation. In our study, we did manual identification of $50$ related work as a QGS~\footnote{Please note that this number includes papers for using network analysis in minable software artifacts too as the plan was to review the application of the network science in software engineering in general. However, we found that splitting the results of the primary studies into two groups: (1) minable human aspects and (2) minable software artifacts would be more appropriate for intensive analysis.} from $8$ related venues by searching their proceedings in the last $10$ years. The first run of the QGS search resulted in a sensitivity of $68\%$ with $16$ missed papers out of the $50$. After inspecting those missed papers, we updated the search terms and reached $100\%$ sensitivity from the second (and the last) run of the QGS search method. The used search queries are in the provided data, and here is a sample of the query used in the IEEE search library:\\
	\begin{scriptsize}
		( "Document Title":software  OR  "Document Title":program OR  "Document Title":evolution OR  "Document Title":maintenance OR  "Document Title":architecture OR  "Document Title":bug OR  "Document Title":maintainability OR  "Document Title":requirements OR  "Document Title":testing OR  "Document Title":clone OR  "Document Title":opensource ) and ("Document Title":graph OR  "Document Title":network OR  "Document Title":tree OR  "Document Title":analysis OR  "Document Title":collaboration OR  "Document Title":call OR  "Document Title":interaction)
	\end{scriptsize}
	\\
	
	\begin{table*}
		\begin{center}
				\begin{scriptsize}
			\begin{tabular}{|c|c|c|c|}
				\hline
				\textbf{Libraries search engine}&	\textbf{Row primary studies}	&\textbf{Initial primary studies} &	\textbf{Primary studies}\\ \hline
				ACM	&1992	&50	&15\\ \hline
				IEEEXplor&	2500&	124	&15\\ \hline
				Springer digital library	&1553	&15	&3\\ \hline
				Elsevier ScienceDirect	&113 &	8	&2\\ \hline
				Wiley open access	&35	&2	&0\\ \hline
				\cellcolor{Black}\color{white}Sum& \cellcolor{Black}\color{white}\textbf{6080}&\cellcolor{Black}\color{white}\textbf{199}&\cellcolor{Black}\color{white}\textbf{35}\\ \hline    
			\end{tabular}%
						\caption{The selection process of the primary studies across the selected libraries. The row primary studies were title-abstract sifted, then the inclusion and exclusion criteria were applied on the initial primary studies resulting in the PSs.}
			\label{tab:selectionprocess}%
		\end{scriptsize}
				\end{center}

	\end{table*}%
	\subsection{Search libraries and In(Ex)clusion criteria}
	\label{subsec:inclusioncriteria}
\begin{figure}
		\begin{center}
			\includegraphics[scale=0.45]{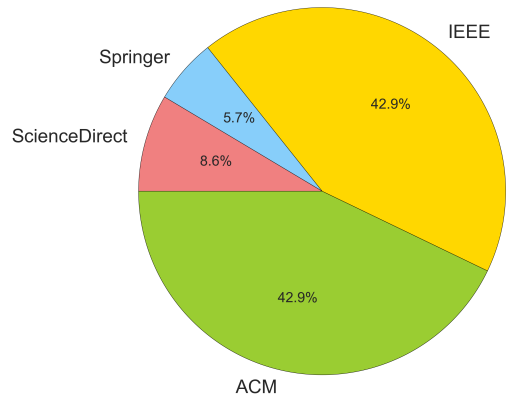}
		\end{center}
		\caption{The percentages of the number of primary studies across the used libraries.}		
		\label{fig:librariespercentages}
	\end{figure}

	We identified the following search libraries as a basis for the search process. The libraries are: (1) The \textit{ACM} digital library, (2) The \textit{IEEExplore}, (3) The \textit{Springer}, and (4) The \textit{ScienceDirect}. The initial studies resulted from the search process was $6080$ that we sifted based on the title and the abstract content. The result was $199$ related initial primary studies that we read in order to apply the inclusion and exclusion criteria. Table~\ref{tab:selectionprocess} and Figure~\ref{fig:librariespercentages} show more details about the number of selected studies in each step across the search libraries. 
	
\begin{figure}
		\begin{center}
			\includegraphics[scale=0.45]{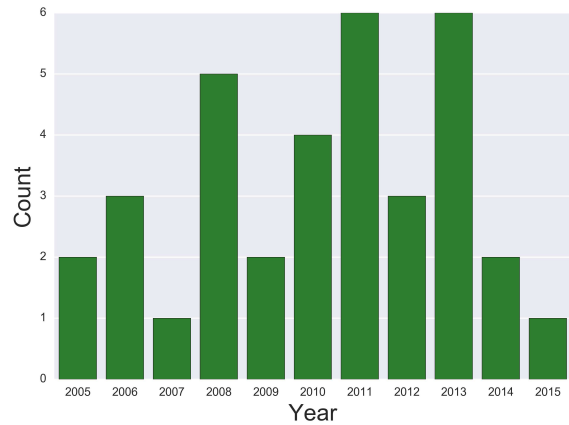}
		\end{center}
		\caption{The distribution of the number of primary studies over years 2005-2015.}		
		\label{fig:years}
	\end{figure}

	Table~\ref{tab:inclusioncriteria} shows the inclusion criteria. Any included study should satisfy all of the inclusion criteria in the table. Table~\ref{tab:exclusioncriteria} shows the exclusion criteria along with the number of excluded studies due to each exclusion criterion.
	\\\\
	\begin{table}
		\center
		\begin{scriptsize}
			\begin{tabular}{|p{2cm}|p{12cm}|}
				\hline 
				\textbf{Inclusion item} & \textbf{Description} \\ 
				\hline 
				$I_1$ & Research that uses network analysis in studying software engineering practices or artifacts or processes.   \\ 
				\hline 
				$I_2$ & Empirical study that provides quantitative results. \\ 
				\hline 
				$I_3$ & It covers minable human aspects and/or minable software artifacts \\ 
				\hline 
				$I_4$ & It is a peer reviewed publications. \\
				\hline 
			\end{tabular}
			\caption{The inclusion criteria applied in this SLR.}
			\label{tab:inclusioncriteria}
		\end{scriptsize}
	\end{table}
	Figure~\ref{fig:years} shows the distribution of the number of primary studies and their publication year. The earlier primary study was published in $2005$, a few years after the emergence of network science papers.
	
\begin{figure}
		\begin{center}
			\includegraphics[scale=0.75]{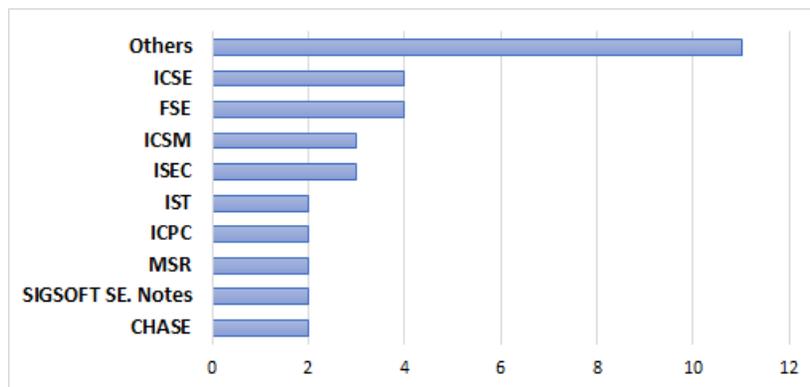}
		\end{center}
		\caption{The distribution of the number of primary studies over the venues.}		
		\label{fig:venues}
	\end{figure}

	\begin{table}
		\center
		\begin{scriptsize}
			\begin{tabular}{|p{2cm}|p{9.9cm}|}
				\hline 
				\textbf{Exclusion item} & \textbf{Description} \\ 
				\hline 
				$E_1$ & Any study that is not in English. ($1$) \\ 
				\hline 
				$E_2$ & Replication studies, invited papers, lessons learned, technical papers, and position papers. ($5$) \\ 
				\hline 
				$E_3$ & Visualization tools, process engineering, and controlled experiment papers. ($14$) \\ 
				\hline 
				$E_4$ & Theoretical models that have not been validated empirically with real systems. ($10$) \\ 
				\hline 
				$E_5$ & Papers that do not cover the minable human aspects in software engineering. ($41$)\\ 
				\hline 
				$E_6$ & Duplicated papers or initial work that has been extended later. ($4$)\\ 
				\hline
				$E_7$ & Papers that only construct networks from data without analysis. ($8$)\\ 
				\hline
			\end{tabular}
			\caption{The exclusion criteria along with the number of excluded studies due to each exclusion criterion.}
			\label{tab:exclusioncriteria}
		\end{scriptsize}
	\end{table}
	After applying the inclusion and exclusion criteria, we got the following primary studies. For the \textit{ACM digital library} and the \textit{IEEE}, we had $15$ studies selected from each. For the \textit{SciencDirect} and the \textit{Springer}, we had $3$ and $2$ primary studies, respectively. The included primary studies in this SLR are shown in Table~\ref{tab:psssummary} along with their venues. Figure~\ref{fig:venues} shows the number of primary studies per venue. The \textit{Others} represents all venues with only one primary study. From Figure~\ref{fig:venues} and from Table~\ref{tab:psssummary}, we noticed that major software engineering conferences like \textit{ICSE} and \textit{FSE} have the highest number of primary studies. Also, from the information in the table and the figure, we see that there is $31$ primary studies published in conference proceedings and the rest, $4$ studies, are published in journals.

	\begin{table*}[htbp]
		\begin{scriptsize}
			\begin{center}
				\begin{tabular}{|c|c|c|l|}
					\hline
					\textbf{Digital library} & \textbf{PS}    & \textbf{Year}  & \multicolumn{1}{c|}{\textbf{Venue\tablefootnote{The full name of the venue can be found in the references.} }} \\
					\hline
					\multirow{7}[30]{*}{ACM} & \cite{Huang:2005:MVH:1082983.1083158} & 2005  & SIGSOFT Softw. Eng. Notes \\
					\cline{2-4}          & \cite{Ohira:2005:ACK:1082983.1083163} & 2005  & SIGSOFT Softw. Eng. Notes \\
					\cline{2-4}          & \cite{Bird:2006:MES:1137983.1138016} & 2006  & MSR \\
					\cline{2-4}          & \cite{Bird:2008:LSS:1453101.1453107} & 2008  & FSE \\
					\cline{2-4}          & \cite{Pohl:2008:DNM:1370114.1370135} & 2008  & CHASE \\
					\cline{2-4}          & \cite{Pinzger:2008:DNP:1453101.1453105} & 2008  & FSE \\
					\cline{2-4}          & \cite{Meneely:2008:PFD:1453101.1453106} & 2008  & FSE \\
					\cline{2-4}          & \cite{Wolf:2009:PBF:1555001.1555017} & 2009  & ICSE \\
					\cline{2-4}          & \cite{Datta2010} & 2010  & ISEC \\
					\cline{2-4}          & \cite{Canfora:2011:SIA:1985441.1985463} & 2011  & MSR \\
					\cline{2-4}          & \cite{Bird:2011:DTM:2025113.2025119} & 2011  & FSE \\
					\cline{2-4}          & \cite{Sureka:2011:USN:1953355.1953381} & 2011  & ISEC \\
					\cline{2-4}          & \cite{Jermakovics:2011:MVD:1984642.1984647} & 2011  & CHASE \\
					\cline{2-4}          & \cite{Kumar:2013:EDS:2442754.2442764} & 2013  & ISEC \\
					\cline{2-4}          & \cite{Panichella:2014:EEC:2597008.2597145} & 2014  & ICPC \\
					\hline
					\hline
					\multirow{7}[30]{*}{IEEE} & \cite{Schwind2008} & 2008  & ECECE \\
					\cline{2-4}          & \cite{Bird2009} & 2009  & ERE \\
					\cline{2-4}          & \cite{Bettenburg2010} & 2010  & ICPC \\
					\cline{2-4}          & \cite{Surian2010} & 2010  & CRE \\
					\cline{2-4}          & \cite{Hong2011} & 2011  & ICSM \\
					\cline{2-4}          & \cite{Meneely2011} & 2011  & ICSE \\
					\cline{2-4}          & \cite{He2012} & 2012  & ICGCC \\
					\cline{2-4}          & \cite{Xuan2012} & 2012  & ICSE \\
					\cline{2-4}          & \cite{Xuan2012a} & 2012  & ICSI \\
					\cline{2-4}          & \cite{Allaho2013} & 2013  & ASONAM \\
					\cline{2-4}          & \cite{Caglayan2013} & 2013  & CHSAE \\
					\cline{2-4}          & \cite{Meng2013} & 2013  & ICSM \\
					\cline{2-4}          & \cite{Zhang2013} & 2013  & ICCSC \\
					\cline{2-4}          & \cite{Bhattacharya2014} & 2014  & ICSME \\
					\cline{2-4}          & \cite{Joblin2015} & 2015  & ICSE \\
					\hline\hline
					\multirow{2}[6]{*}{ScienceDirect} & \cite{Sowe2006} & 2006  & EOSSD \\
					\cline{2-4}          & \cite{Xu2006} & 2006  & IST \\
					\cline{2-4}          & \cite{Toral2010} & 2010  & IST \\
					\hline\hline
					\multirow{2}[4]{*}{Springer} & \cite{Gao2007} & 2007  & OSDAI \\
					\cline{2-4}          & \cite{Dittrich2013} & 2013  & Complex networks \\
					\hline
				\end{tabular}
			\end{center}
			\caption{A summary of the identified primary studies in this SLR.}
			\label{tab:psssummary}%
		\end{scriptsize}
	\end{table*}%

	\subsection{Quality assessment}
	\label{subsec:qualityassessment}
	Performing quality assessment is important in SLRs in order to get insights regarding the quality of the included PSs and to perform a proper analysis for them. We followed the method presented in~\cite{Dyba2007} for the quality assessment task. We restricted our quality assessment to the research quality of the PSs as the other aspects were covered in the data extraction in order to answer the research questions in Section~\ref{subsec:researchquestions}. We did not apply any exclusion based on the quality of the PSs as this will limit our ability to answer the research question properly.
	
	\begin{table}
		\center
		\begin{scriptsize}
			\begin{tabular}{|r|p{3cm}|p{10.5cm}|}
				\hline 
				\textbf{Code} & \textbf{Addresses} &\textbf{Description} \\ 
				\hline 
				$QA_1$ & Research Design& The study provides a clear research design that includes research goals, hypotheses and other aspects of research design. \\ 
				\hline 
				$QA_2$ &  Data Collection& The study provides a clear information about the data used in the analysis and how it was collected and validated. \\ 
				\hline 
				$QA_3$ & Threats to Validity (or Limitation)& The study provides the threats that affect the validation of the study including internal, external, and construction threats. \\ 
				\hline 
				$QA_4$ &  Empirical Results Interpretation & The study provides sufficient and thorough interpretation of the results. \\ 
				\hline 
				$QA_5$ &  Reflections on SE& The study reflects and links the results to the studied aspects of software engineering.\\ 
				\hline 
				$QA_6$ & Reproducibility&  The study provides sufficient information to reproduce the experiments. E.g., links to the datasets, the steps in details...etc.\\ 
				\hline
			\end{tabular}
			\caption{The quality assessment items. The evaluation of each item can take a descriptive value from Explicit, Implicit, or None with numeric values $1$, $0.5$, or $0$. }
			\label{tab:qualityassessment}
		\end{scriptsize}
	\end{table}
	Based on the defined quality criteria shown in Table~\ref{tab:qualityassessment}, we got the quality scores for all of the primary studies as shown in Table~\ref{tab:qualityassessmentscors}. The average quality score for all primary studies is $3.9$ out of $5$. The retrieved primary studies from the $IEEE$ library have, on average, higher quality than the studies from the other libraries. The average quality scores for the primary studies retrieved from the \textit{ACM}, \textit{ScienceDirect}, and the Springer libraries are $3.9$, $3.2$, and $2$, respectively. The results of quality assessment show that there are $5$ PSs with quality score $6$, which is the maximum, out of the $35$ primary studies. The lowest quality score was $0.5$ for only one primary study. We also calculated the average score for each quality aspect shown in Table~\ref{tab:qualityassessment}. The table shows that the average value for the quality assessment item $QA_6$ was the lowest among the other quality assessment items, while the highest average value was for $QA_4$. We found a $0.4$ positive correlation, using the $R^2$ coefficient, between the quality of the primary studies published in the \textit{IEEE} over time. This correlation was insignificant for the other libraries.
	
\begin{figure}
		\begin{center}
			\includegraphics[scale=0.5]{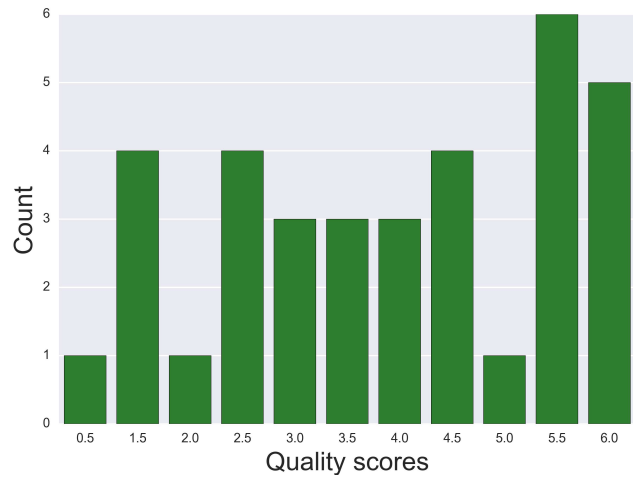}
		\end{center}
		\caption{The distribution of the number of primary studies over the quality scores.}		
		\label{fig:scores}
	\end{figure}

	Figure~\ref{fig:scores} shows the distribution of the quality scores of the primary studies. 
	
	\begin{table*}[h]
		\begin{center}
			\begin{scriptsize}
				\begin{tabular}{|c|c|c|c|c|c|c|c|}
					\hline
					\multicolumn{1}{|c|}{\multirow{2}[4]{*}{\textbf{PS}}} & \multicolumn{6}{c|}{\textbf{Quality Aspect}}          & \multicolumn{1}{c|}{\multirow{2}[4]{*}{\textbf{Quality score}}} \\
					\cline{2-7}          & \multicolumn{1}{l|}{$QA_1$} & \multicolumn{1}{l|}{$QA_2$} & \multicolumn{1}{l|}{$QA_3$} & \multicolumn{1}{l|}{$QA_4$} & \multicolumn{1}{l|}{$QA_5$} & \multicolumn{1}{l|}{$QA_6$} &  \\
					\hline
					\cite{Wolf:2009:PBF:1555001.1555017} & 1     & 1     & 1     & 1     & 1     & 0.5   & \cellcolor{gray}\color{white}5.5 \\
					\hline
					\cite{Sureka:2011:USN:1953355.1953381} & 0.5   & 0     & 0     & 0     & 0     & 0     & \cellcolor{gray}\color{white}0.5 \\
					\hline
					\cite{Pohl:2008:DNM:1370114.1370135} & 0.5   & 0.5   & 0     & 0.5   & 0     & 0     & \cellcolor{gray}\color{white}1.5 \\
					\hline
					\cite{Pinzger:2008:DNP:1453101.1453105} & 1     & 0.5   & 1     & 1     & 1     & 0     & \cellcolor{gray}\color{white}4.5 \\
					\hline
					\cite{Panichella:2014:EEC:2597008.2597145} & 1     & 1     & 1     & 1     & 1     & 1     & \cellcolor{gray}\color{white}6 \\
					\hline
					\cite{Ohira:2005:ACK:1082983.1083163} & 0     & 1     & 0     & 0.5   & 1     & 0     & \cellcolor{gray}\color{white}2.5 \\
					\hline
					\cite{Meneely:2008:PFD:1453101.1453106} & 0     & 1     & 1     & 0.5   & 0.5   & 1     & \cellcolor{gray}\color{white}4 \\
					\hline
					\cite{Kumar:2013:EDS:2442754.2442764} & 0.5   & 1     & 1     & 1     & 1     & 0     & \cellcolor{gray}\color{white}4.5 \\
					\hline
					\cite{Jermakovics:2011:MVD:1984642.1984647} & 0.5   & 1     & 1     & 0.5   & 0     & 0     & \cellcolor{gray}\color{white}3 \\
					\hline
					\cite{Huang:2005:MVH:1082983.1083158} & 0.5   & 0     & 0     & 0.5   & 0.5   & 0     & \cellcolor{gray}\color{white}1.5 \\
					\hline
					\cite{Datta2010} & 1     & 1     & 0     & 1     & 1     & 0.5   & \cellcolor{gray}\color{white}4.5 \\
					\hline
					\cite{Canfora:2011:SIA:1985441.1985463} & 1     & 1     & 1     & 1     & 1     & 1     & \cellcolor{gray}\color{white}6 \\
					\hline
					\cite{Bird:2011:DTM:2025113.2025119} & 1     & 1     & 0     & 1     & 1     & 0.5   & \cellcolor{gray}\color{white}4.5 \\
					\hline
					\cite{Bird:2008:LSS:1453101.1453107} & 1     & 1     & 1     & 1     & 1     & 0.5   & \cellcolor{gray}\color{white}5.5 \\
					\hline
					\cite{Bird:2006:MES:1137983.1138016} & 1     & 1     & 0     & 1     & 0.5   & 0     & \cellcolor{gray}\color{white}3.5 \\
					\hline
					\cite{Zhang2013} & 1     & 1     & 1     & 1     & 1     & 1     & \cellcolor{gray}\color{white}6 \\
					\hline
					\cite{Xuan2012a} & 0     & 1     & 0     & 1     & 1     & 0.5   & \cellcolor{gray}\color{white}3.5 \\
					\hline
					\cite{Xuan2012} & 1     & 1     & 1     & 1     & 1     & 0.5   & \cellcolor{gray}\color{white}5.5 \\
					\hline
					\cite{Surian2010} & 1     & 1     & 0     & 0     & 0     & 0.5   & \cellcolor{gray}\color{white}2.5 \\
					\hline
					\cite{Schwind2008} & 0.5   & 0     & 0     & 1     & 0     & 0     & \cellcolor{gray}\color{white}1.5 \\
					\hline
					\cite{Meng2013} & 0.5   & 0.5   & 0     & 1     & 0.5   & 0.5   & \cellcolor{gray}\color{white}3 \\
					\hline
					\cite{Meneely2011} & 1     & 1     & 1     & 1     & 1     & 1     & \cellcolor{gray}\color{white}6 \\
					\hline
					\cite{Joblin2015} & 1     & 1     & 1     & 1     & 1     & 1     & \cellcolor{gray}\color{white}6 \\
					\hline
					\cite{Hong2011} & 1     & 0     & 1     & 1     & 0.5   & 0     & \cellcolor{gray}\color{white}3.5 \\
					\hline
					\cite{He2012} & 1     & 0.5   & 1     & 1     & 1     & 0.5   & \cellcolor{gray}\color{white}5 \\
					\hline
					\cite{Caglayan2013} & 1     & 0     & 1     & 1     & 1     & 0     & \cellcolor{gray}\color{white}4 \\
					\hline
					\cite{Bird2009} & 1     & 0.5   & 0     & 0.5   & 0.5   & 0     & \cellcolor{gray}\color{white}2.5 \\
					\hline
					\cite{Bhattacharya2014} & 1     & 1     & 1     & 1     & 1     & 0.5   & \cellcolor{gray}\color{white}5.5 \\
					\hline
					\cite{Bettenburg2010} & 0.5   & 1     & 1     & 1     & 1     & 1     & \cellcolor{gray}\color{white}5.5 \\
					\hline
					\cite{Allaho2013} & 1     & 1     & 0     & 1     & 0.5   & 0.5   & \cellcolor{gray}\color{white}4 \\
					\hline
					\cite{Toral2010} & 0.5   & 0.5   & 0     & 1     & 1     & 0     & \cellcolor{gray}\color{white}3 \\
					\hline
					\cite{Sowe2006} & 0.5   & 1     & 1     & 1     & 1     & 1     & \cellcolor{gray}\color{white}5.5 \\
					\hline
					\cite{Xu2006} & 0     & 0.5   & 0     & 1     & 0     & 0.5   & \cellcolor{gray}\color{white}2 \\
					\hline
					\cite{Gao2007} & 0     & 0     & 0     & 1     & 0.5   & 0     & \cellcolor{gray}\color{white}1.5 \\
					\hline
					\cite{Dittrich2013} & 0     & 1     & 0.5   & 0.5   & 0     & 0.5   & \cellcolor{gray}\color{white}2.5 \\
					\hline
					Avg.& \cellcolor{gray}\color{white}0.69	&\cellcolor{gray}\color{white}0.7	&\cellcolor{gray}\color{white}0.53	&\cellcolor{gray}\color{white}0.84 &\cellcolor{gray}\color{white} 0.7	&\cellcolor{gray}\color{white}0.4	&\cellcolor{gray}\color{white} 3.9\\\hline
				\end{tabular}%
								\caption{The quality assessment of the primary studies.}
				\label{tab:qualityassessmentscors}%
			\end{scriptsize}
		\end{center}
	\end{table*}%

	\subsection{Data extraction and synthesis}
	\label{subsec:dataextraction}
	Data synthesis for heterogeneous studies is always hard, particularity when the primary studies cover the same aspect with different measurements and present the results in different ways. Also, the primary studies in most cases use a mixed-methods for the analysis that makes a universal comparison impossible. Thus, we followed a meta-ethnography method~\cite{britten2002}, as advised by Silva et al. in~\cite{da2013}, which translates the quantitative findings of the PSs into comparable features that can be used for a proper analysis. Thus, we invested a lot of time during the execution of this SLR in designing a good set of features to be extracted from the primary studies in order to answer the research questions. Table~\ref{tab:extractionfields} shows the set of features extracted from each primary study. Features in category "A" are documentary features that capture the general goal of each primary study. For space limitation, we will discuss few of them in the following. Feature "A1" in this review has the same value (i.e., \textit{Human Aspects} that is the main topic of this review). Feature "A2" represents the SE context that the proposed network analysis method serves, for example, how can we do bug assignment by analyzing the developers collaboration network?. Feature "A3" captures the network being used in each PS. Features of category "B" address the network analysis aspects of the PSs, and features of category "C" address the other aspects. Features "B1" to "B6" address questions about the constructed network type. This is an important step towards understating how valid the constructed networks are, which is covered in features "B7" to "B10". Feature "B7" reflects how good an edge is. Real edges are real connections between the nodes. A proxy edge is used when the real edge is hard to obtain. For example, changing a file, a module, or any software artifact may not yield a real relationship. Instead, we call this interaction a proxy. An example of a real edge is the edges constructed in the communication email networks, in which the edge and its direction are explicitly found in reality. Having clear idea about the constructed network, we can check the validity of the used measures and their meaningfulness. Features "B11" to "B13" address the used measures in each PS. Feature "B12" is important and it reflects whether the meaning of the used measures is sufficiently explained or not, and if the explanation exists, whether it is meaningful or not. The feature "B14" tests whether the results in each PS are validated or not.\\
	The extraction features in Table~\ref{tab:extractionfields} are designed to be mapped to the research questions in Section~\ref{subsec:researchquestions}. Extraction features "B1" to "B10" give us rich information that enable obtaining strong evidence regarding the quality of the constructed networks addressed in \textbf{RQ1}. Features "B11" to "B13" answer \textbf{RQ2}, while the feature "B14" is mapped to \textbf{RQ3}. Features "C1" to "C4" answer \textbf{RQ4}. The answers of the first four research questions enable us to answer \textbf{RQ5}.

	\begin{table*}[h]
		\begin{center}
		\begin{small}
			\begin{tabular}{|c|p{13.8cm}|}
				\hline
				\textbf{Key} & \multicolumn{1}{c|}{\textbf{Description}} \\
				\hline
				A1    & Human aspects or software artifact aspects \\
				\hline
				A2    & The addressed context. Examples: bug tracking and team formation.  \\
				\hline
				A3    & What is the constructed network? Examples: Developer-Module network and social networks of developers. \\
				\hline\hline
				B1    & What is a node? \\
				\hline
				B2    & What is an edge? \\
				\hline
				B3    & Is the network multiplex (Multilayer)? \\
				\hline
				B4    & Is the network bipartite (2-mode networks)? \\
				\hline
				B5    & Is the network dynamic or static?\\
				\hline
				B6    & Is the network directed or undirected ?\\
				\hline
				B7    & Edge quality.  \\
				\hline
				B8    & Is the used network aggregated from multiplex interactions? \\
				\hline
				B9    & Is the used network aggregated from longitudinal networks? \\
				\hline
				B10   & Are edge weights (like collaboration intensity/frequency) considered? \\
				\hline
				B11   & What are the used measures? \\
				\hline
				B12   & Are the meaning of the used measures explained? \\
				\hline
				B13   & Does the work contribute new measures, tools, models, and algorithms that are based on network analysis? \\
				\hline
				B14   & Does the work use any Null Model to validate the results? \\
				\hline\hline
				C1    & Data source project. Open source and/or industry \\
				\hline
				C2    & The analyzed project(s) \\
				\hline
				C3    & Link the results and interpret them in the context of SE, i.e., A2  \\
				\hline
				C4    & Practical implications, for examples, lessons or advice for practitioners  \\
				\hline
			\end{tabular}%
		\end{small}
		\end{center}
		\caption{The extraction features of the SLR.}
				\label{tab:extractionfields}
	\end{table*}%
	
	\section{Results}
	\label{sec:results}
	In this section, we will provide the synthesized results, while we provide an evidence-based answers to the research questions in Section~\ref{sec:discussion}.
	\subsection{From SE to a network (Fields A)}
	\label{subsec:fromsetonetwork}
\begin{figure}
		\begin{center}
			\includegraphics[width=0.8\textwidth]{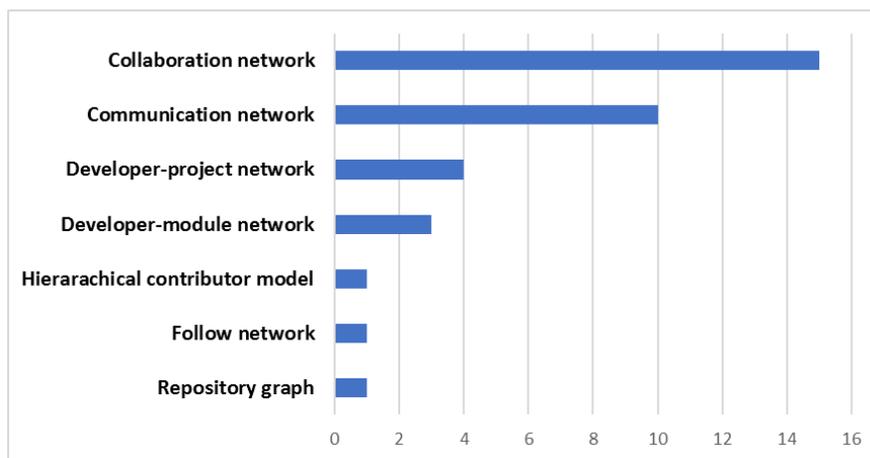}
		\end{center}
		\caption{The constructed networks in the PSs and their frequencies.}		
		\label{fig:networks}
	\end{figure}

	In this section, we provide the results for the extraction features "A1" to "A3". As per the inclusion and exclusion criteria described in Section~\ref{subsec:inclusioncriteria}, all of the included PSs were handling \textit{Human Aspects} of software engineering. Thus, the $35$ PSs had that value in that field. Feature $A2$ had different values that handled different contexts of SE. There are $8$ PSs that handled \textit{Bug} related issues like bug assignment, bug tracking and bug prediction. Other PSs addressed different contexts like \textit{Build Failure}, \textit{Code Ownership}, \textit{Team Organization}, \textit{Failure Prediction}, \textit{Code Quality}, and \textit{Team Collaboration}.\\
	Figure~\ref{fig:networks} shows the used networks in the PSs. The most used network, in $15$ PS, was the \textit{Collaboration Network} where the nodes were developers and an edge appeared between any two developers if they worked together on the same file, line of code, module, or project. If the collaboration network was based on more than one type of interaction between the developers, then this was a one-mode projection from a multiplex bipartite network. The \textit{Communication network} was constructed by modeling the developers as nodes and the edges were communication messages between developers. The \textit{Developer-project network} was a bipartite network where the two sets $V_L$ and $V_R$ were the developers and the projects, respectively, and the edges represented developers taking part in a project. Similarly, the \textit{Developer-module network} was a bipartite network between developers and the modules. Those two networks were found in $7$ primary studies, ~\cite{Pinzger:2008:DNP:1453101.1453105,Ohira:2005:ACK:1082983.1083163,Bird:2011:DTM:2025113.2025119,Bird2009,Xu2006,Gao2007,Dittrich2013}. The \textit{Follow network} was the social network of the Github platform where a developer could explicitly follow another one, like Twitter. The other two networks, the \textit{Hierarchical} and \textit{Repository} networks were devised and contributed networks of the works in~\cite{Allaho2013,Meng2013}, respectively.\\
	\subsection{The constructed networks, their validity, and measures validation (Features B)}
	\label{subsec:thevalidityofnetwork}
\begin{figure}
		\begin{center}
			\includegraphics[width=0.8\textwidth]{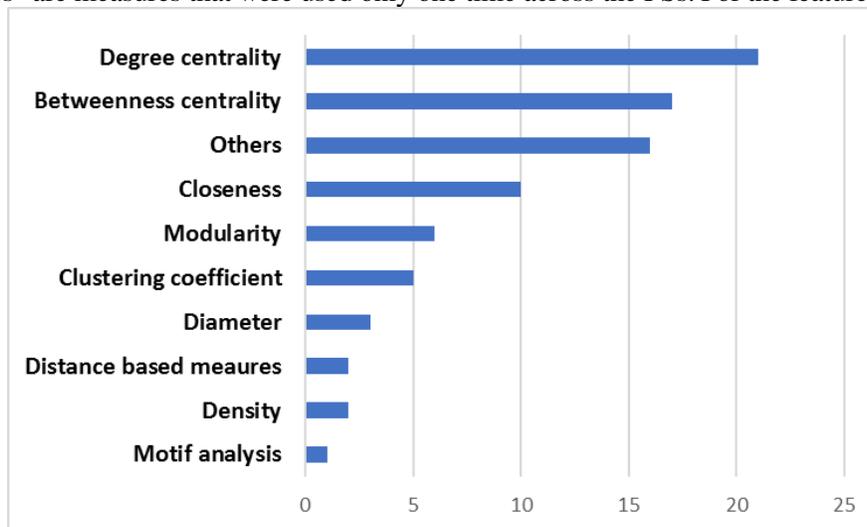}
		\end{center}
		\caption{The used measures in the PSs and their frequencies.}		
		\label{fig:measures}
	\end{figure}

	Not surprisingly, the nodes in the majority of the networks of the PSs were \textit{Contributors}, which means any human who contributed to a software repository, like a developer, a tester, an architect, etc. An exception to this was the work in~\cite{Meng2013,Bird2009,Xu2006,Gao2007,Dittrich2013} where the nodes were two sets of a bipartite network and the analysis was done on the constructed bipartite network, not on a one-mode projected network.\\
	For feature "B3", there were $3$ PSs, ~\cite{Bird:2011:DTM:2025113.2025119,Zhang2013,Bhattacharya2014}, where the constructed network was a multiplex network. Additionally, only $5$ PSs, \cite{Pohl:2008:DNM:1370114.1370135,Kumar:2013:EDS:2442754.2442764,Xuan2012a,He2012,Toral2010}, considered the temporality of a network and provided a \textit{dynamic} network. The rest of the primary studies considered a \textit{static} network. The PSs that considered directed networks were $13$ PSs, ~\cite{Wolf:2009:PBF:1555001.1555017,Canfora:2011:SIA:1985441.1985463,Bird:2006:MES:1137983.1138016,Zhang2013,Xuan2012,Schwind2008,Meng2013,Joblin2015,Bird2009,Bhattacharya2014,Allaho2013,Toral2010,Sowe2006} and the rest PSs considered an undirected network. We found that $60\%$ of the PSs, $21$ PSs, used a \textit{proxy} edge. Only $5$ PSs, ~\cite{Wolf:2009:PBF:1555001.1555017,Pohl:2008:DNM:1370114.1370135,Panichella:2014:EEC:2597008.2597145,Bhattacharya2014,Xu2006}, used a multiplex representation of the network. The rest used a unipelx network. Nearly $89\%$ of the PSs aggregated the networks over time and used a single static network, only $4$ primary studies, ~\cite{Pohl:2008:DNM:1370114.1370135,Kumar:2013:EDS:2442754.2442764,Xuan2012a,He2012} considered different networks of the same nodes over time and performed analysis based on this situation. For feature "B10", only $13$ PSs considered the weights of the edges, while the other $22$ PSs did not consider the edge weights.\\
	The used measures in the PSs are shown in Figure~\ref{fig:measures}. The figure shows that the \textit{Degree} and the \textit{Betweenness} centrality measures were dominant among the used measures with $21$ and $17$ times used in the PSs, respectively. Motif analysis was used only once in the work~\cite{Surian2010}. The measures in "Others" are measures that were used only one time across the PSs.
	For the feature "B12", we found that $11$ PSs did not provide sufficient definition and explanation for the used measures, and only $9$ PSs, ~\cite{Pinzger:2008:DNP:1453101.1453105,Panichella:2014:EEC:2597008.2597145,Bird:2011:DTM:2025113.2025119,Zhang2013,Xuan2012a,Meng2013,Joblin2015,Caglayan2013,Bhattacharya2014}, contributed (devised) new measures and(or) models for the constructed networks. Surprisingly, we found only $14\%$, $5$ PSs out of the $35$, that used a \textit{Null model} to validate the results. The rest of the PSs did not use any statistical validation of the results.\\
	\subsection{Data, results, and interpretation (Feature C)}
	\label{subsec:dataresultsandinterpretation}
	The used data set in the PSs was distributed as follows: (1) $6$ PSs used data from industry, (2) $27$ PSs used data from open source platforms, and (3) $2$ PSs, ~\cite{Bird2009,Jermakovics:2011:MVD:1984642.1984647}, used both industry and open source data in the experiments. From the open source project, the projects \textit{Mozilla}, \textit{Apache}, \textit{Eclipse}, and \textit{Linux} were used much often. For the industry projects, the \textit{MS-Vista binaries} were used $2$ times by the same author in~\cite{Bird:2011:DTM:2025113.2025119,Pinzger:2008:DNP:1453101.1453105}. The MS-Vista binaries was used also in addition to an open source project (i.e., the \textit{Eclipse}) in the work~\cite{Bird2009}. Two PSs did not provide information about the used industrial data~\cite{Datta2010,Caglayan2013}. The projects Eclipse, Firefox, and Apache were used the most in the PSs with frequencies $7$, $6,$ and $4$ times, respectively. The answer to feature "C3" found as follows. Only $34\%$ of the PSs, $12$ PSs, provided an \textit{Explicit} interpretation of the results in the context of the studied context of SE, feature "A2". Other $14$ studies provided \textit{Implicit} interpretation of the results, which means it was not clear in the primary study how the context in "A2" was related to the results \textit{directly}. The rest of the results, $9$ PSs, did not provide any information about connecting the results to feature "A2".\\
	The PSs that provided \textit{Explicit} lessons and guidelines for the practitioners based on the results of the study were $5$ studies only, ~\cite{Wolf:2009:PBF:1555001.1555017,Pinzger:2008:DNP:1453101.1453105,Panichella:2014:EEC:2597008.2597145,Bird:2011:DTM:2025113.2025119,Bird:2008:LSS:1453101.1453107}. Also, $4$ studies ~\cite{Ohira:2005:ACK:1082983.1083163,Canfora:2011:SIA:1985441.1985463,Bettenburg2010,Toral2010} provided \textit{Implicit} lessons and guidelines for the practitioners, and the rest did not provide any information.
	
	\section{Discussions}
	\label{sec:discussion}
	In this section, we discuss the results presented in Section~\ref{sec:results}, and provide answers to the research questions in Section~\ref{subsec:researchquestions}.
	
	\subsection{Discussion around RQ1:}
	\label{subsec:answerrq1}
	To answer this question, we need to test the quality of the constructed network. To that end, we will use the results of features "B1" to "B10" to answer this question. The quality of the network stems from the quality of its components (i.e., the nodes and the edges) and on the other aspects we mentioned in the definition section like being weighted, multiplex, or bipartite. Based on the results, we found that $60\%$ of the PSs used a proxy edge. A proxy edge is obtained by doing one (or more) of the following:
	\begin{itemize}
		\item Classical OMP: which was described in the Section~\ref{sec:definitions}. The PSs
		~\cite{Dittrich2013,Gao2007,Xu2006,Bhattacharya2014,Ohira:2005:ACK:1082983.1083163,Toral2010} did classical OMP, which does not respect the edge weights. That means, the links between any two developers, for example, in the network who collaborated $1$ time or $1000$ times are treated equally. The resulted network is not informative and a lot of measure, like the degree centrality, do not provide meaningful results for the projected networks if the weights are ignored.
		\item Multiplex(space) aggregation: In this type of aggregation, like PSs~\cite{Bhattacharya2014,Bird:2008:LSS:1453101.1453107,Toral2010,Wolf:2009:PBF:1555001.1555017,Panichella:2014:EEC:2597008.2597145,Xu2006}, the edges are aggregated from different types of relationships(interactions). This results in a dense disguised network, in which the link do not reflect any special interaction that can be meaningfully quantified. Edge aggregation from different types of interactions is an oversimplification of reality, which yields wrong conclusions as shown by Cardillo et al.~\cite{cardillo2013}.
		\item Temporal(time) aggregation: In this type of aggregation, like the \\PSs~\cite{Ohira:2005:ACK:1082983.1083163,Meneely:2008:PFD:1453101.1453106,Jermakovics:2011:MVD:1984642.1984647,Huang:2005:MVH:1082983.1083158,Datta2010,Bird:2008:LSS:1453101.1453107,Schwind2008,Meng2013,Meneely2011,Joblin2015,Hong2011,Bhattacharya2014,Bettenburg2010,Toral2010,Sowe2006,Xu2006,Gao2007,Dittrich2013}, the edges are aggregated over time into a single network.
	\end{itemize}
	The problem of the previous classical OMP and the aggregations is that it generates a very dense graphs, where every node is almost connected to every other node in the network. This renders many measures, like the degree centrality, not of much benefit.
	Some researchers were aware of this problem and included this as a threat to validity in their work, like the PS~\cite{Jermakovics:2011:MVD:1984642.1984647}.
	Additionally, the bipartite networks, the dynamic networks, and the multiplex networks have their special characteristic that differs from the unipartite static network. Thus, converting any non-unipartite, non-static, or multiplex network into a unipartite static network should be done with extreme care if required, and generally we recommend keeping the network in its nature. Additionally, the collaboration intensity (or frequency) was not considered when it should be. There are $22$ PSs that did not consider the weights of the edges in the graph where the weights are crucial.
	Thus, our answer to the \textit{RQ1} is: \textit{In $60\%$ of the PSs, the constructed networks are not of sufficient quality to be used as a model for the studied context. Accordingly, we believe that rigorous validation need to take place for the subsequent analysis results of those PSs.}
	
	To address the problems found in the PSs for this research question, we provide here some pointers to extend the space of the used methods. For example, the work by Zweig~\cite{zweig2011} provides a systematic method to do a proper one mode projection, the works by Kivel{\"a} et al.~\cite{kivela2014multilayer} and by Boccaletti et al.~\cite{boccaletti2014} provide a good resource for the methods, the models, and the measures that can be used to analyze the multiplex networks, and the seminal work by Holme and Saram{\"a}ki~\cite{holme2012temporal} provide frameworks, measures, and methods for the temporal networks (dynamic networks).
	\subsection{Discussion around RQ2:}
	\label{subsec:answerrq2}
	To answer this question, we will provide a discussion on the used measures in the PSs, and extensively talk about one of them, which is the betweenness centrality that was used too much. Borgatti~\cite{borgatti2005} showed that many of the centrality measures embraced unstated assumptions that once were not satisfied, the results of the measures could not be reliable and interpretable. For space limitation, we will discuss the betweenness centrality that was used $16$ times in the PSs. This centrality measure was introduced by Freeman~\cite{1978freeman1centrality} and implicitly assumed that: (1) there is a process going on top of the network, (2) the process is based on the shortest paths, (3) the process takes place among all pairs of nodes with the same frequency, and (4) the process is sequential~\cite{Zweig:2014}. Thus, any network where those assumptions are not met can not benefit from the betweenness centrality as a measure for identifying central nodes. Based on this information and the information provided in Section~\ref{subsec:answerrq1}, none of the networks in the PSs is suitable to apply the betweenness centrality on it. Any obtained results based on this measure and these networks are not meaningful. Similar situations to this are also found with other measures. For example, the diameter of the network and the closeness centrality do not give an informative value in the analysis of a social interaction. Those measures are informative in networks where the cost of establishing an edge is high, like adding new street in the street networks and adding new DNS server in the internet network. However, for communication and social networks where an edge is a sent email, and for the collaboration networks where the edge is a repository fork or a comment added to the code, the diameter, for example, is just meaningless and does not provide a value to the analysis.
	In addition, only $9$ PSs, ~\cite{Pinzger:2008:DNP:1453101.1453105,Panichella:2014:EEC:2597008.2597145,Bird:2011:DTM:2025113.2025119,Zhang2013,Xuan2012a,Meng2013,Joblin2015,Caglayan2013,Bhattacharya2014}, contributed a new model or a new measure based on the specifics of the used networks, the other $26$ PSs only used an existing measure in the literature. Also, $11$ primary studies did not provide an explanation and interpretation over the used measure. Thus, our answer to \textit{RQ2} is: \textit{In at least $25$ PSs, the used measures were not suitable and were not reliable. Also, in $11$ PSs, the provided measures were not sufficiently explained in the context of used network. Thus, the results of those PSs need a proper justification for their usage validity.}
	To tackle this issue, we recommend designing models and devising measures for each constructed network if the existing methods and measures are not suitable. In the primary studies covered in this SLR, there are $9$ PSs,~\cite{Pinzger:2008:DNP:1453101.1453105,Panichella:2014:EEC:2597008.2597145,Bird:2011:DTM:2025113.2025119,Zhang2013,Xuan2012a,Meng2013,Joblin2015,Caglayan2013,Bhattacharya2014}, that provided new measures and models specifically for the constructed network that respected the nature of the modeled system.
	
	\subsection{Discussion around RQ3:}
	\label{subsec:answerrq3}
	Classically, the random graphs null models are used to validate the significance of the results obtained from networks. For our PSs, only $14\%$ of them, $5$ PSs,\cite{Bird:2011:DTM:2025113.2025119,Bird:2008:LSS:1453101.1453107,Joblin2015,Hong2011,Caglayan2013}, out of the $35$, used a \textit{Null model} to validate the results. The rest of the PSs used no validation method at all. We noticed that those studies used the random graphs null model to validate the quality of the found clusters (community). Thus, our answer to the \textit{RQ3} is: \textit{The results of $86\%$ of the PSs in this SLR were not validated at all. Accordingly, the reported results of these studies may not be reliable.}
	To overcome this issue, we provide here some pointers for the null models used to validate the significance of the results. Newman~\cite{newman2002random} provided a description of using random graphs as a null model. The work by Schlauch et al.~\cite{schlauch2015different} provided a comprehensive comparison between different null models.
	\subsection{Discussion around RQ4:}
	\label{subsec:answerrq4}
	We noticed that $37\%$ of the PSs, $3$ out of $8$, used industry data sets and $24\%$ of the PSs, $7$ out of $29$, used open source data sets provided \textit{Explicit} lessons and guidelines for the practitioners. Based on that and on the results provided in Section~\ref{subsec:dataresultsandinterpretation}, our answer to the \textit{RQ4} is: \textit{Translating the results of the analysis into the studied context of SE is poor. Also, the number of PSs that provided lessons and guidelines for practitioners is very limited.}
	\subsection{Discussion around RQ5:}
	\label{subsec:answerrq5}
	\textit{RQ5} has been partially answered after answering \textit{RQ1}, \textit{RQ2}, \textit{RQ3}, and \textit{RQ4}.
	Here we elaborate more on the answer for \textit{RQ5} based on the results in Section~\ref{sec:results} and based on the answers to the research questions in Sections~\ref{subsec:answerrq1} to~\ref{subsec:answerrq4}. Another contribution in this paper is the following items that may help the researchers to further extend the body of the work in mining software repositories using network science.
	\begin{enumerate}
		\item \textit{Temporal measures:} As we saw in the previous sections, the constructed networks were in many times aggregated over time and the static measures were applied, which weakens the validity of the results or simply turns them not meaningful~\cite{kim2012temporal}. Thus, we recommend using the dynamic networks and their corresponding measures as provided by Kim et al.~\cite{kim2012temporal} and by Holme et al.~\cite{holme2012temporal}. We strongly think that utilizing temporal networks as a model to understand the dynamics of collaboration and its effect would give more actionable insights.
		\item \textit{Weighted networks:} The collaboration between developers and the communication between them can hardly be imagined without considering the intensity (or frequency) of this collaboration or communication, which is a rich information that should be utilized. The work by Opsahl~\cite{opsahl2010node} and by Newman~\cite{newman2004analysis} provide methods for bipartite and weighted networks and their corresponding centrality measures.
		\item \textit{Community detection}: The PSs that handle community detection were based on the work of Newman~\cite{newman2004detecting}. A recent work by Ahn et al.~\cite{ahn2010link} showed promising results for finding structural properties in networks based on the link communities. This method could provide more robust team organizations with the collaborative development.
		\item \textit{Link prediction and assessment}: The prediction of collaboration or communication between developers was, surprisngly, not found in the PSs. In network science there is a plethora of work in link prediction started by Liben-Nowell and Kleinberg~\cite{liben2007link}. The link prediction is an active area in network science and can be utilized in mining software repositories, especially when incorporating external information~\cite{abufouda2014,abufouda2015,abufouda2017}. An application of the link prediction in this context could be built as recommendation system for bringing developers to projects in open source.
	\end{enumerate}

	\subsection{Closing thoughts}
	\label{subsec:closingthoughts}
	\begin{itemize}
		\item \textit{The collaboration networks}: in the PSs were projected from different interactions. They were proxy networks from the one mode projection. In most cases, we could not retrieve any information about the procedures done by the researchers in order to get the OMP.
		\item \textit{The communication networks}: they were in some cases projected from bipartite network of thread-developer network, where a thread was a message that was sent to all developers and any one could reply it. This results in a very dense network that was not really a communication network.
		\item \textit{Dynamic vs Static}: Should be considered carefully. While static networks give some insights regarding the modeled network, a dynamic analysis of a network gives richer insights regarding the process that is being done over this network. 
		\item \textit{Bipartite networks}: should be utilized as bipartite networks when possible. This requires extra measures and more models that are context dependent.
		\item \textit{In-applicability of some measures}: not every measure can provide meaningful results for every network. A measure should be first understood well in order to use it and get useful insights.
	\end{itemize}
	
	\subsection{Threats to validity}
	\label{subsec:threatstovalidity}
	
	\begin{itemize}
		\item \textit{Completeness}: The completeness of any SLR is hard to attain. We described the steps we followed to cover as relevant studies as possible.
		\item \textit{Reproducibility}: We incorporated the QGS in order to guarantee the reproducibility of the results in the SLR. All of the data used for this review is available upon request.
	\end{itemize}
	
	\section{Conclusion}
	\label{sec:conclusion}
	In this work, we have presented a systematic literature review to identify and evaluate the use of network science as a tool to understand the collaboration of developers in software engineering. We followed the most rigor steps we know in performing a systematic review. We identified $35$ primary studies, assessed their quality, and extracted the data from them. The data extracted from the primary studies was used to answer $5$ research questions. Our answers showed that the primary studies used networks that lack the required quality, they used in most cases unsuitable measures, and in most cases they do not validate the results. We provided some pointers that can be utilized in future research works in order to do a good network analysis with actionable insights.

	\bibliography{references_combined}

\end{document}